\documentclass[pre,twocolumn,noshowpacs,a4paper]{revtex4}
\usepackage{graphicx}

\newcommand{\mymat}[1]{\mathbf{#1}}

\newcommand{\M}{\mymat{M}}
\newcommand{\Mid}{\M_{\mathrm{id}}}
\newcommand{\Mex}{\M_{\mathrm{ex}}}

\newcommand{\eqref}[1]{(\ref{#1})}
\newcommand{\citen}{\cite}

\begin{document}

\title{Flory-Huggins theory for the solubility of
heterogeneously-modified polymers}

\author{Patrick B. Warren}
\affiliation{Unilever R\&D Port Sunlight, Bebington, Wirral, CH63 3JW, UK.}

\date{April 5, 2007}

\begin{abstract}
Many water soluble polymers are chemically modified versions of
insoluble base materials such as cellulose.  A Flory-Huggins model is
solved to determine the effects of heterogeneity in modification on
the solubility of such polymers.  It is found that heterogeneity leads
to decreased solubility, with the effect increasing with increasing
blockiness.  In the limit of extreme blockiness, the nature of the
phase coexistence crosses over to a polymer-polymer demixing
transition.  Some consequences are discussed for the synthesis of
partially modified polymers, and the experimental characterisation of
such systems.
\end{abstract}

\maketitle

Many water-soluble polymers are made by chemically modifying insoluble
base materials such as starches and gums, for example a wide class of
water-soluble polymers are obtained from cellulose \cite{gumbook,
Reuben}.  It is often possible to vary the degree of modification of
the base polymer to obtain water soluble polymers with, in principle,
continuously variable properties.  A basic characteristic of these
polymers is their solubility, but given the essentially stochastic
nature of the chemical modification step, what is the effect of
heterogeneity in modification on the solubility of the resulting
materials?

In the present paper, this question is approached from a theoretical
point of view by setting up a Flory-Huggins model for the phase
behaviour of a polymer-solvent mixture \cite{Flory}, where the
polymers have a random degree of modification.  In this approach, the
issue of solubility is translated into the problem of determing the
phase coexistence between a dissolved aqueous phase and an undissolved
(water-poor) phase.  The solubility is then formally given by the
polymer concentration in the aqueous phase.  Determination of the full
phase behaviour for a multicomponent Flory-Huggins theory is an
onerous task though, and a simpler approach is to examine the spinodal
stability of the system, which can be taken to be representative of
the full phase behaviour.  This is the approach taken in the present
paper.  It is arguably more insightful than a full calculation of the
phase behaviour since closed-form analytic expressions can be obtained
for the spinodal stability limit.  The approach taken is similar to
models for the phase behaviour of random block copolymer melts which
have been developed in the past \cite{FM, FML, NOdlCC}.  There has
been rather little work though on random copolymers which also include
a solvent, apart from a brief example described by Sollich {\it et al}
\cite{SWC}.

In the present model, it is supposed that the system comprises a large
number of species of polymers $i$ with differing degrees of
modification $0<\alpha_i<1$ and concentrations $\rho_i$.  For
simplicity, length polydispersity is neglected, and all the polymers
are assumed to have the same number $N$ of segments.  The system is
then described by the following (mean field) Flory-Huggins free energy
density,
\begin{equation}
\textstyle
f = \sum_i\rho_i\log\rho_i
+(1-\phi)\log(1-\phi)+\chi(\phi-\eta)(1-\phi),
\label{eq:f}
\end{equation}
where $\phi$ is the total polymer segment concentration and $\eta$ is
the concentration of chemically modified segments, given respectively
by $\phi=N\sum_i\rho_i$ and $\eta=N\sum_i\rho_i\alpha_i$.  The first
term in Eq.~\eqref{eq:f} is the ideal free energy of mixing.  The
second term is the usual Flory-Huggins configurational chain entropy.
The third term is the free energy cost of the unmodified polymer
segments at a concentration $\phi-\eta$ coming into contact with
solvent (water) at a concentration $1-\phi$.  Typically one expects
$\chi>1/2$ for this interaction, to represent the repulsion between
unmodified segments and water which leads to phase separation of
unmodified polymers.  To keep the model simple, this is the only
$\chi$-parameter that is retained in the problem.

Eq.~\eqref{eq:f} has the structure of a \emph{moment free energy},
since the excess free energy, comprising the second and third terms,
only depends on $\phi$ and $\eta$ which are \emph{moment densities}.
Such a system can be analysed using the methods developed by Sollich
and coworkers \cite{SWC, SC-PRL, W-PRL, W-EPL}.  In particular,
Ref.~\citen{W-EPL} describes how the spinodal stability conditions for
systems with an excess free energy can be expressed in terms of
moment densities, generalising various truncation theorems obtained by
earlier workers \cite{IG, Hendriks}.  I now summarise the relevant
results, translated into terms suitable for the present problem.  Let
us consider such a system with a free energy $f=\sum_i\rho_i\log\rho_i
+ f^{\mathrm{(ex)}}(\phi^{(1)} \dots \phi^{(n)})$, where the excess
free energy depends on moment densities of the form $\phi^{(r)} =
\sum_i \rho_i\, w^{(r)}_i$ ($r=1\dots n$), with the $w_i^{(r)}$ being
species-dependent weights.  The fundamental idea is that the moment
densities can be treated as effective species concentrations.  In
particular, it can be proved that spinodal stability corresponds to
the positive-definiteness of the matrix $\M$ of second partial
derivatives of the free energy with respect to the moment densities.
In Ref.~\citen{W-EPL} it is shown that $\M = \Mid + \Mex$ where
$(\Mid^{-1})_{rs} = \sum_i \rho_i \,w_i^{(r)} w_i^{(s)}$ and
$(\Mex)_{rs} = {\partial ^2\!f^{\mathrm{(ex)}}} \!/
{\partial\phi^{(r)} \partial\phi^{(s)}}$.  The limit of spinodal
stability is given by $\det\M=0$.  This condition usually corresponds
to the vanishing of a single eigenvalue of $\M$, with an eigenvector
$\Delta\phi^{(s)}$ that satisfies $\sum_s(\M)_{rs}\Delta\phi^{(s)}=0$.
It is shown in Ref.~\citen{W-EPL} that the spinodal instability
direction in the space of species concentrations is given by
$\Delta\rho_i = \sum_{rs} \rho_i\, w_i^{(r)} (\Mid)_{rs}
\Delta\phi^{(s)}$.  

For the present problem, there are two moment densities $\phi$ and
$\eta$, defined respectively with $w_i^{(1)}=N$ (a constant) and
$w_i^{(2)} = N\alpha_i$ (the number of modified groups on the $i$th
species).  Application of the above theory to Eq.~\eqref{eq:f} leads to
\begin{equation}
\Mid^{-1} = N^2
\left(\begin{array}{cc}
\sum_i\rho_i & \sum_i\rho_i\alpha_i\\[6pt]
\sum_i\rho_i\alpha_i & \sum_i\rho_i\alpha_i^2
\end{array}\right)\label{eq:mat1}
\end{equation}
and
\begin{equation}
\Mex = \left(\begin{array}{cc}
({1-\phi})^{-1}-2\chi & \chi\\[6pt]
\chi & 0
\end{array}\right).\label{eq:mat3}
\end{equation}
After some algebra the condition $\det\M=0$ reduces to
\begin{equation}
\frac{1}{N\phi}+\frac{1}{1-\phi}
-2\chi(1-\langle\alpha\rangle)
-\chi^2N\phi(\langle\alpha^2\rangle-\langle\alpha\rangle)
=0,\label{eq:poly}
\end{equation}
where
\begin{equation}
\textstyle
\langle\alpha\rangle = {\sum_i\rho_i\alpha_i}\,/\,{\sum_i\rho_i},\quad
\langle\alpha^2\rangle = {\sum_i\rho_i\alpha_i^2}\,/\,{\sum_i\rho_i}.
\label{eq:aa2}
\end{equation}
I emphasise that, despite being remarkably simple, Eq.~\eqref{eq:poly}
is exact.

One already reaches a significant conclusion from this.  The first
three terms in Eq.~\eqref{eq:poly} are what one would expect from
standard Flory-Huggins theory \cite{Flory}, with an effective
$\chi$-parameter given by the product of the original $\chi$-parameter
and the fraction $1-\langle\alpha\rangle$ of unmodified segments.
These terms therefore take account of the \emph{mean} degree of
modification.  The final term in Eq.~\eqref{eq:poly} is a correction
due to the heterogeneity.  Since the variance $\langle\alpha^2\rangle
- \langle\alpha\rangle^2$ is positive, this term is always negative.
The effect is that heterogeneity in modification \emph{reduces} the
solubility, over and above what would be expected from the mean degree
of modification.

To make further progress, it is convenient to specify a model for the
distribution of the $\alpha_i$.  In particular, such a model can be
used to examine the effect of blockiness in modification which is
expected to play an important role.  In previous work on random block
copolymers \cite{FM, FML}, a Markov model was used to characterise the
correlations between different kinds of segments.  Whilst such a model
may be appropriate for the stochastic nature of the synthetic route
for such random block copolymers, as discussed below it is probably
not appropriate in the present case.  I therefore consider instead a
very simple model for the heterogeneity in which the modified segments
occur in blocks of size $M$, where $1<M<N$.  In this model, it is
supposed that each block has an equal probability $p$ of being
modified, and there are no further correlations.  Then, for any
particular species, $\alpha_i = ({1}/{N})\sum_{j=1}^{N/M} M
\epsilon_{ij}$ where $j$ labels the blocks, and $\epsilon_{ij}$ is
zero or one with probability $1-p$ and $p$ respectively.  Thus the
$\alpha_i$ are drawn from scaled binomial distribution, with
\begin{equation}
\langle\alpha\rangle = p,\quad
\langle\alpha^2\rangle - \langle\alpha\rangle = ({M}/{N})\,p(1-p).
\end{equation}
Eq.~\eqref{eq:poly} becomes
\begin{equation}
\frac{1}{N\phi}+\frac{1}{1-\phi}
-2\chi(1-p)
-\chi^2M\phi p(1-p)
=0.\label{eq:polyM}
\end{equation}
This is a quadratic equation for $\chi$ and the appropriate root is
\begin{equation}
\chi=\frac{1}
{M\phi p}
\Bigl[\Bigl\{
1+\frac{M\phi p}{1-p}
\Bigl(\frac{1}{N\phi}+\frac{1}{1-\phi}\Bigr)
\Bigr\}^{1/2}-1\Bigr].
\label{eq:fullM}
\end{equation}
I now examine the consequences of this result.

The formal limit $M\to0$ corresponds to a vanishing variance and a
completely uniform distribution of modified segments, as though each
monomer has undergone an identical fractional modification by a
fraction $p$, rather than being modified or not with probability $p$
and $1-p$.  As noted already above, this limit corresponds to simple
Flory-Huggins theory with an effective $\chi$-parameter equal to
$\chi(1-p)$.  For large $N$, this indicates the absence of phase
separation for $\chi(1-p) < 1/2$ or $p > 1-1/(2\chi)$.

\begin{figure}
\includegraphics{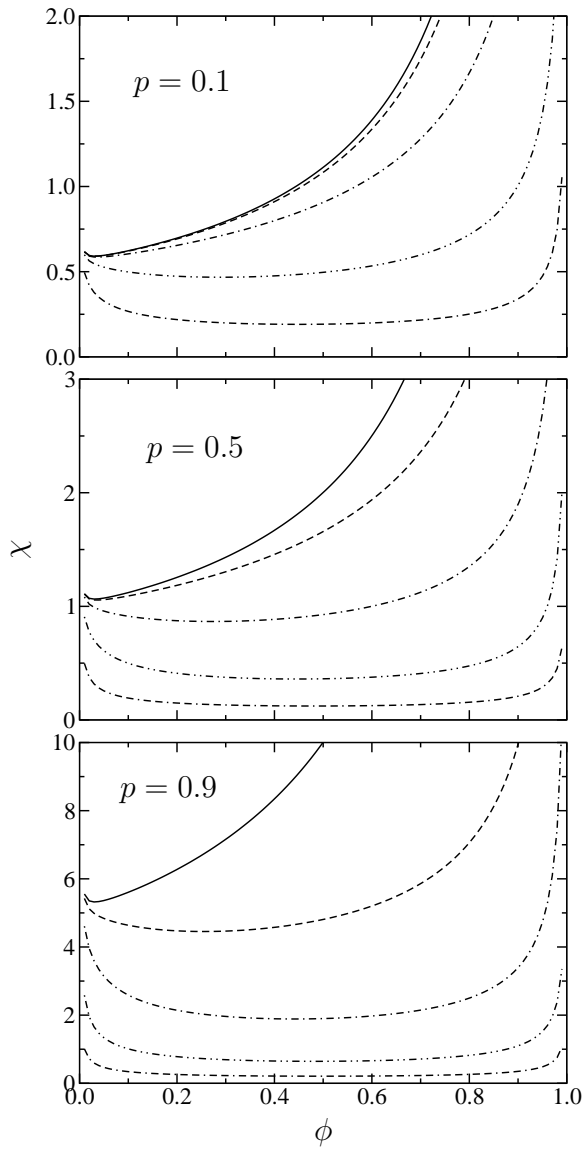}
\caption[]{Spinodal curves calculated from Eq.~\eqref{eq:fullM} for
polymers of length $N=10^3$, for three values of the mean degree of
modification $p$, and for block sizes $M\to0$ (uniform limit, solid
line), $M=1$ (dashed line), $M=10$ (dash-dot line), $M=100$
(dash-dot-dot line) and $M=10^3$ (dash-dash-dot line).  The system is
spinodally unstable above the indicated curves.  Note the change in
shape of the $M=1$ curves: for $p=0.1$ and 0.5 the minimum is at
$\phi\to0$, whereas for $p=0.9$ the minimum is at $\phi\approx
0.25$.\label{fig:spin}}
\end{figure}

\begin{figure}
\includegraphics{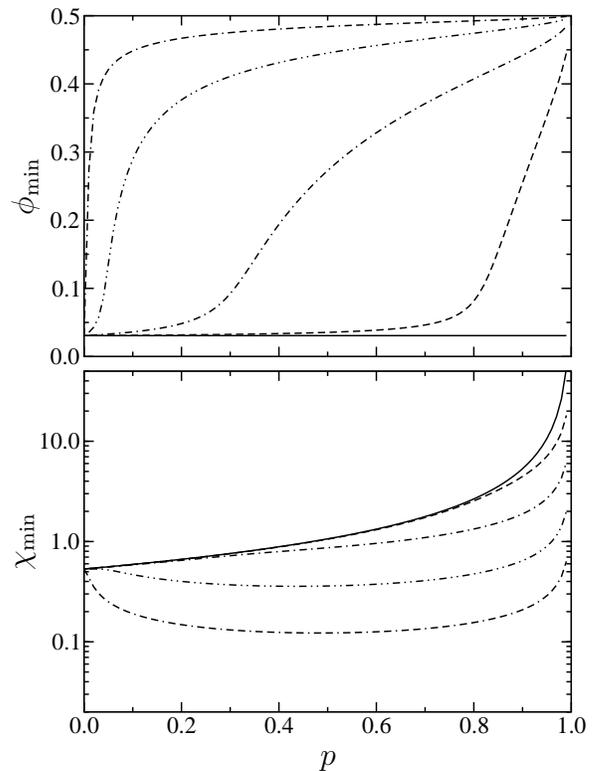}
\caption[]{The location of the numerically determined minimum of the
spinodal curves from Eq.~\eqref{eq:fullM} is plotted as a function of
$p$, for polymers of length $N=10^3$ and block sizes $M\to0$ (uniform
limit, solid line), $M=1$ (dashed line), $M=10$ (dash-dot line),
$M=100$ (dash-dot-dot line) and $M=10^3$ (dash-dash-dot line).  For
$M=1$ (dashed line) the upper plot shows clearly that
$\phi_{\mathrm{min}}\approx N^{-1/2}\approx0.03$ only holds for $p \alt
4/5=0.8$.\label{fig:min}}
\end{figure}

Now let us consider Eq.~\eqref{eq:fullM} for block size $M=1$.  In
this case, individual segments are modified randomly with no
correlations.  For $M=1$ and large $N$ in Eq.~\eqref{eq:fullM}, there
are two behaviours depending on the value of $p$.  For $p<4/5$, there
is an absence of phase separation for $\chi(1-p) < 1/2$, just as
for the $M\to0$ limit.  For $4/5<p<1$, the behaviour is more
complicated.  To be precise, the location of the minimum value of the
$\chi(\phi)$ spinodal shifts from $\phi_{\mathrm{min}} \sim N^{-1/2}$
for $p<4/5$ to a non-vanishing $0<\phi_{\mathrm{min}}<1$ for $p>4/5$
(it is the examination of Eq.~\eqref{eq:fullM} in the limit $\phi\sim
N^{-1/2}$ that gives the cross over point $p=4/5$).  The change in
behaviour can be seen for the $M=1$ curves (dashed lines) in
Fig.~\ref{fig:spin} and is shown explicitly in the upper plot of
Fig.~\ref{fig:min}.

Let us next consider the limit of extreme blockiness $M=N$.  This
limit is strikingly different from the $M=1$ case.  For large $N$ and
$p>0$, one can show that there is an absence of phase separation only
for $\chi \sqrt{Np(1-p)} < 2$.  In the large $N$ limit, this
inequality is always violated, indicating that the system
\emph{always} has a tendency to undergo phase separation in the limit
of extreme blockiness.  Since the unmodified polymer system itself
only phase separates for $\chi > 1/2$, this suggests that the phase
separation has the nature of a polymer-polymer demixing transition
rather than a solvent-driven phase separation.  This insight is
confirmed by analysis of the spinodal instability direction below.

For large $N$ and general $M$ in Eq.~\eqref{eq:fullM}, one would
expect that the above two cases represent the two classes of
behaviour.  In the first case $M\ll N$ and the behaviour is similar to
the $M=1$ limit where individual segments are randomly modified.  In
the second case, $M\propto N$ and the behaviour is similar to the
$M=N$ limit of extreme blockiness.  Fig.~\ref{fig:spin} shows typical
spinodal curves calculated from Eq.~\eqref{eq:fullM} for various
values of $p$ and $M$.  The location of the minimum
$(\phi_{\mathrm{min}}, \chi_{\mathrm{min}})$ of the spinodal curves
can be numerically determined, and Fig.~\ref{fig:min} shows how this
depends on $p$.

The results show firstly that for $M\ll N$, increasing $p$ leads to
increasing solubility as the value of $\chi$ required to reach the
spinodal instability is increased.  Moreover, a decrease in solubility
between a uniform model ($M\to0$) with no heterogeneity, and a model
with fine-grained blockiness ($M=1$), is apparent.  The major
effect arises as $M\to N$ though, where the tendency for phase
separation is greatly enhanced.

The above analysis is augmented considering the spinodal instability
direction associated with the spinodal stability limit which can
provide a useful mechanistic insight.  As explained above, the
spinodal instability direction is characterised by the eigenvector
that corresponds to the vanishing eigenvalue responsible for the
vanishing spinodal determinant.  For the present problem, from
Eqs.~\eqref{eq:mat1}--\eqref{eq:mat3}, one finds the instability
direction is characterised by
\begin{equation}
{\Delta\eta}\,/\,{\Delta\phi} = 
\langle\alpha\rangle
-\chi N\phi(\langle\alpha^2\rangle-\langle\alpha\rangle^2)
\end{equation}
The corresponding spinodal instability direction in the space of
species concentrations is
\begin{equation}
\begin{array}{ll}
\displaystyle
\frac{\Delta\rho_i}{\rho_i} &
\displaystyle
=
\frac{\langle\alpha^2\rangle\Delta\phi-\langle\alpha\rangle\Delta\eta
+\alpha_i(\Delta\eta-\langle\alpha\rangle\Delta\phi)}
{\phi(\langle\alpha^2\rangle-\langle\alpha\rangle^2)}\\[12pt]
&
\displaystyle
= \frac{\Delta\phi}{\phi}\Bigl(1 + \chi N\phi 
(\langle\alpha\rangle-\alpha_i)\Bigr)
\end{array}
\label{eq:sid}
\end{equation}
where the second line follows by inserting the result for the ratio
$\Delta\eta/\Delta\phi$.  These results should be evaluated on the
spinodal.  They are all exact, for an arbitrary distribution of
$\alpha_i$.

For the instability direction to lie along a pure dilution line, one
should have $\Delta\rho_i/\rho_i$ independent of species $i$.  One can
conclude that this only happens if $\Delta\eta/\Delta\phi =
\langle\alpha\rangle$, in other words if the variance
$\langle\alpha^2\rangle - \langle\alpha\rangle^2$ vanishes.  In such a
case, the phase transition is purely associative, or solvent-driven,
meaning that the compositions of the coexisting phases remain the same
($\Delta\rho_i/\Delta\phi=\rho_i/\phi$).

If one specialises to the model of blockiness described above by
inserting the value of $\chi$ corresponding to the spinodal stability
limit, the instability direction becomes
\begin{equation}
\frac{\Delta\eta}{p\,\Delta\phi}=1-\frac{1-p}{p}
\Bigl[\Bigl\{
1+\frac{M\phi p}{1-p}
\Bigl(\frac{1}{N\phi}+\frac{1}{1-\phi}\Bigr)
\Bigr\}^{1/2}-1\Bigr].
\end{equation}
This confirms that the spinodal instability lies along a dilution
line ($\Delta\eta/\Delta\phi = \langle\alpha\rangle = p$) only in the
limit $M\to0$ which formally corresponds to a vanishing variance.  For
$M=1$ (and $M\ll N$ in general) the phase transition has a mixed
character.  The interesting case occurs when $M=N$ (or $M\propto N$ in
general) for which $\Delta\eta/\Delta\phi\sim (-)N^{1/2}$ in the limit
of large $N$.  One can write this as $\Delta\phi/\Delta\eta\to0$ as
$N\to\infty$.  This shows that the phase transition tends towards
being purely segregative, meaning that the overall polymer
concentration in coexisting phases remains the same ($\Delta\phi=0$).
This confirms the suggestion above, that in the limit of extreme
blockiness, the system tends towards a segregative polymer-polymer
demixing transition.

Let us now try to draw some conclusions.  The main effect of
randomness is to reduce the solubility of partially-modified polymers
beyond what would be expected from the mean degree of modification.
The extent to which this occurs depends on the blockiness in
substitution.  For fine-grained blockiness, the phase behaviour is
expected to be similar to a system for which there is no randomness,
albeit with a somewhat reduced solubility.  For coarse-grained
blockiness, where the block size is comparable to the polymer length,
the nature of the phase transition changes to a polymer-polymer
demixing transition.  In this situation, one expects that the modified
polymers (being almost fully modified) will partition into the aqueous
phase, leaving the unmodified polymers behind.

The reason for considering the two extreme kinds of blockiness is now
clearer: namely one can envisage two different mechanisms of chemical
modification (this is the reason why a Markov model for the
distribution of modified segments has not been used).  Fine-grained
blockiness would arise if monomers are equally accessible to the
modifying agent, irrespective of their surroudings.  If this cannot be
achieved in a one-step process (for the reason described below) it
could perhaps be achieved in a two-step process, by fully modifying
the polymers then removing a random fraction of the derivative groups.
Extreme blockiness on the scale of the polymer chain itself would
arise if the modifying agent was present only in the aqueous phase,
and as such only able to access polymer which had already been
solubilised.  This would lead to a mixture of polymers which were
either fully modified, or remained unmodified and insoluble.  The
process of modification of insoluble polymers could still be initiated
because the modifying agent is able to access the tiny proportion of
the insoluble polymer segments which lie at the interface between the
insoluble and aqueous phases.  Experimentally, confirmation of the
scenario of extreme blockiness would be given by measuring the mean
degree of modification for the \emph{dissolved} polymers.  One should
find that this is much in excess of the apparent mean degree of
modification.

In the calculation, the major effect arises from inter-chain rather
than intra-chain heterogeneities.  The model is not sophisticated
enough to take account of the solution structures such as micelles or
mesophases that could form for blocky polymers with block sizes
$M\gg1$ but still $M<N$ (for example, diblock copolymers).  Such
polymers would be expected to have greater solubilities than would be
predicted from the Flory-Huggins theory since the hydrophobic groups
can be buried in micelles or other solution structures.  The present
theory could be extended to discuss these inhomogeneous situations
using a Landau approach developed for random block copolymers
\cite{FM, FML, NOdlCC}.  For the mechanistic routes discussed above
though, it is difficult to envisage that polymers with intermediate
block sizes could arise very easily.  I therefore expect that the
general conclusions will remain.

Finally I note that in principle the above model for the phase
behaviour could be combined with a model for the chemical modification
reaction, to obtain a theory for reaction-induced solubility.
However, one needs to take great care to capture the kinetics
correctly \cite{BC}.

I thank Nigel Clarke for a critical reading of the manuscript.


\begin{thebibliography}{13}
\expandafter\ifx\csname natexlab\endcsname\relax\def\natexlab#1{#1}\fi
\expandafter\ifx\csname bibnamefont\endcsname\relax
  \def\bibnamefont#1{#1}\fi
\expandafter\ifx\csname bibfnamefont\endcsname\relax
  \def\bibfnamefont#1{#1}\fi
\expandafter\ifx\csname citenamefont\endcsname\relax
  \def\citenamefont#1{#1}\fi
\expandafter\ifx\csname url\endcsname\relax
  \def\url#1{\texttt{#1}}\fi
\expandafter\ifx\csname urlprefix\endcsname\relax\def\urlprefix{URL }\fi
\providecommand{\bibinfo}[2]{#2}
\providecommand{\eprint}[2][]{\url{#2}}

\bibitem[{\citenamefont{Davidson}(1980)}]{gumbook}
\bibinfo{author}{\bibfnamefont{R.~L.} \bibnamefont{Davidson}},
  \emph{\bibinfo{title}{Handbook of water-soluble gums and resins}}
  (\bibinfo{publisher}{McGraw-Hill}, \bibinfo{address}{New York},
  \bibinfo{year}{1980}).

\bibitem[{\citenamefont{Rueben}(1984)}]{Reuben}
\bibinfo{author}{\bibfnamefont{J.}~\bibnamefont{Rueben}},
  \bibinfo{journal}{Macromol.} \textbf{\bibinfo{volume}{17}},
  \bibinfo{pages}{156} (\bibinfo{year}{1984}).

\bibitem[{\citenamefont{Flory}(1953)}]{Flory}
\bibinfo{author}{\bibfnamefont{P.~J.} \bibnamefont{Flory}},
  \emph{\bibinfo{title}{Principles of polymer chemistry}}
  (\bibinfo{publisher}{Cornell University Press}, \bibinfo{address}{Ithaca, New
  York}, \bibinfo{year}{1953}).

\bibitem[{\citenamefont{Fredrickson and Milner}(1991)}]{FM}
\bibinfo{author}{\bibfnamefont{G.~H.} \bibnamefont{Fredrickson}}
  \bibnamefont{and} \bibinfo{author}{\bibfnamefont{S.~T.}
  \bibnamefont{Milner}}, \bibinfo{journal}{Phys. Rev. Lett.}
  \textbf{\bibinfo{volume}{67}}, \bibinfo{pages}{835} (\bibinfo{year}{1991}).

\bibitem[{\citenamefont{Fredrickson et~al.}(1992)\citenamefont{Fredrickson,
  Milner, and Leibler}}]{FML}
\bibinfo{author}{\bibfnamefont{G.~H.} \bibnamefont{Fredrickson}},
  \bibinfo{author}{\bibfnamefont{S.~T.} \bibnamefont{Milner}},
  \bibnamefont{and} \bibinfo{author}{\bibfnamefont{L.}~\bibnamefont{Leibler}},
  \bibinfo{journal}{Macromolecules} \textbf{\bibinfo{volume}{25}},
  \bibinfo{pages}{6341} (\bibinfo{year}{1992}).

\bibitem[{\citenamefont{Nesariker et~al.}(1993)\citenamefont{Nesariker, {Olvera
  de la Cruz}, and Crist}}]{NOdlCC}
\bibinfo{author}{\bibfnamefont{A.}~\bibnamefont{Nesariker}},
  \bibinfo{author}{\bibfnamefont{M.}~\bibnamefont{{Olvera de la Cruz}}},
  \bibnamefont{and} \bibinfo{author}{\bibfnamefont{B.}~\bibnamefont{Crist}},
  \bibinfo{journal}{J. Chem. Phys.} \textbf{\bibinfo{volume}{98}},
  \bibinfo{pages}{7385} (\bibinfo{year}{1993}).

\bibitem[{\citenamefont{Sollich et~al.}(2001)\citenamefont{Sollich, Warren, and
  Cates}}]{SWC}
\bibinfo{author}{\bibfnamefont{P.}~\bibnamefont{Sollich}},
  \bibinfo{author}{\bibfnamefont{P.~B.} \bibnamefont{Warren}},
  \bibnamefont{and} \bibinfo{author}{\bibfnamefont{M.~E.} \bibnamefont{Cates}},
  \bibinfo{journal}{Adv. Chem. Phys.} \textbf{\bibinfo{volume}{116}},
  \bibinfo{pages}{265} (\bibinfo{year}{2001}).

\bibitem[{\citenamefont{Sollich and Cates}(1998)}]{SC-PRL}
\bibinfo{author}{\bibfnamefont{P.}~\bibnamefont{Sollich}} \bibnamefont{and}
  \bibinfo{author}{\bibfnamefont{M.~E.} \bibnamefont{Cates}},
  \bibinfo{journal}{Phys. Rev. Lett.} \textbf{\bibinfo{volume}{80}},
  \bibinfo{pages}{1365} (\bibinfo{year}{1998}).

\bibitem[{\citenamefont{Warren}(1998)}]{W-PRL}
\bibinfo{author}{\bibfnamefont{P.~B.} \bibnamefont{Warren}},
  \bibinfo{journal}{Phys. Rev. Lett.} \textbf{\bibinfo{volume}{80}},
  \bibinfo{pages}{1369} (\bibinfo{year}{1998}).

\bibitem[{\citenamefont{Warren}(1999)}]{W-EPL}
\bibinfo{author}{\bibfnamefont{P.~B.} \bibnamefont{Warren}},
  \bibinfo{journal}{Europhys. Lett.} \textbf{\bibinfo{volume}{46}},
  \bibinfo{pages}{295} (\bibinfo{year}{1999}).

\bibitem[{\citenamefont{Irvine and Gordon}(1981)}]{IG}
\bibinfo{author}{\bibfnamefont{P.}~\bibnamefont{Irvine}} \bibnamefont{and}
  \bibinfo{author}{\bibfnamefont{M.}~\bibnamefont{Gordon}},
  \bibinfo{journal}{Proc. R. Soc. Lond. A} \textbf{\bibinfo{volume}{375}},
  \bibinfo{pages}{397} (\bibinfo{year}{1981}).

\bibitem[{\citenamefont{Hendriks}(1988)}]{Hendriks}
\bibinfo{author}{\bibfnamefont{E.~M.} \bibnamefont{Hendriks}},
  \bibinfo{journal}{Ind. Eng. Chem. Res.} \textbf{\bibinfo{volume}{27}},
  \bibinfo{pages}{1728} (\bibinfo{year}{1988}).

\bibitem[{\citenamefont{Buxton and Clarke}(2005)}]{BC}
\bibinfo{author}{\bibfnamefont{G.~A.} \bibnamefont{Buxton}} \bibnamefont{and}
  \bibinfo{author}{\bibfnamefont{N.}~\bibnamefont{Clarke}},
  \bibinfo{journal}{Macromolecules} \textbf{\bibinfo{volume}{38}},
  \bibinfo{pages}{8929} (\bibinfo{year}{2005}).

\end{thebibliography}

\end{document}